\documentstyle[12pt]{article}
\begin{document}
\baselineskip .3in
\begin{center}
{\large{\bf {A Study on the Structure Functions and the Radius of the Nucleons }}} \\
\vskip .2in
A. BHATTACHARYA$^{*}$, A.SAGARI, B. CHAKRABARTI$^{+}$ and S. MANI \\
\vskip .1in
Department of Physics, Jadavpur University \\
Calcutta 700032, India. \\
\end{center}
\vskip .3in {\centerline{\bf Abstract}} We have investigated the
properties of the structure functions of the nucleons in the
context of the statistical model using only the radius parameter
of the respective nucleons. The radius of the neutron has been
estimated and is found to be 0.8fm. It is interesting to observe
that the proton radius 0.865fm which exactly equals to the most
accepted charge radius of the proton and which in turn
characterizes proton as a meso object, yields reasonable results
for the structure function and its properties. GT sum rule has
been investigated with some interesting conclusions.

 \vskip .3in

\noindent {\bf PACS No.s:} 12.39.-x, 12.39.Jh, 12.39.Pn, 14.20Dh,

\noindent {\bf Keywords:} Structure functions, GT Sum Rule,radius
of nucleon, meso object, fractal dimension.
 \vskip .3in $*$
  e-mail: pampa@phys.jdvu.ac.in
  \vskip .3in
  $+$ Permanent Address: Department of Physics, Jogamaya Devi
  College, Calcutta, India.

 \vskip.3in

\newpage

There have been several attempts towards the understanding and
estimating of the structure function of the nucleons [1]. It is
well known that the measurement of nucleon structure function is
important for the understanding of the internal structure of
nucleons. The most current interest in structure function is due
to the violation of GT Sum rule which indicates the sea quark
asymmetry inside the nucleons.  The difference between the neutron
and the proton structure functions is observed to be sensitive to
the u and d quark distribution.  NMC [2] result describes the
deviation of the GT Sum rule [3] in nucleons. It has been
suggested that the pion cloud contribution plays a crucial role in
the long range structure of the nucleons. Zamani [4] has
investigated the $F_{2}$ structure function of nucleons using
meson cloud in a light cone frame. They have investigated GT Sum
rule violation and argued that the contribution from the meson
cloud is small which indicates the other sources of violation.
Farrar et al [5] have investigated pion nucleon structure function
near x$\longrightarrow$1 in coloured quark and vector gluon model.
Osipenko et al [6] have investigated the contribution of leading
twist moment of neutron structure function. In a subsequent
analysis [7] they have investigated proton $F_{2}$ structure
function in the resonance region and investigated the evolution of
its moments. Broadhurst et al [8]have shown that at the two loop
level the GT sum rule is suppressed by a factor
$\frac{1}{N_{c}^{2}}$. Cheng et al [9] have investigated the
contribution of the strange quark content of the nucleon to the
deviation of the GT sum rule in the context of the chiral quark
model.
 In the present work we have investigated the $F_{2}$
structure function of the proton and neutron in the context of the
statistical model [10]. In The framework of the model the
structure functions can be investigated via the radius parameter
of the respective nucleons. It has been observed that the neutron
radius 0.80fm and proton radius 0.865fm reproduces the structure
functions and its properties reasonably well within the other
theoretical and experimental estimates. The violation of the GT
sum rule has been investigated and the violation has been
suggested to be the manifestation of the difference of the spacial
extension of the nucleons which in turn may be related to the
fractal behaviour of the hadrons [11].

\newpage
$\bf{The}$ $\bf{Model}$

 In statistical model [10] the probability
density for a nucleon is obtained as:
 \begin{equation}
 |\Psi(r)|^{2} = \frac{315}{64\pi
 r_{0}^{9/2}}(r_{0}-r)^{3/2}\theta(r_{0}-r)
 \end{equation}
  where $r_{0}$ is the radius parameter of the nucleon and $\theta$ is usual step function.
  The momentum space wave function $\psi(k)$ is derived from the
 Fourier transform of the wave function $\psi(r)$ as:
 \begin{equation}
 \psi(k) = \frac{c}{k}\int_{0}^{\infty}r\psi(r)dr.sinkr
 \end{equation}
 where c is appropriate normalisation constant.The normalised
 momentum space wave function with (1) is obtained as:
 \begin{equation}
 \psi(k) = 2\sqrt{3\pi r_{0}} k^{-1}j_{1}(kr_{0})
 \end{equation}
 It is to be noted that $\psi(k)$ depends only on the corresponding
 size parameter of the nucleon.In our subsequent analysis we would
 investigate the nucleon structure function with the above
 $\psi(k)$ as an input.

$\bf {Radius}$ $\bf{of}$ $\bf{ Neutron}$ and $\bf{Proton:}$

 The radius of the neutron has been derived by adjusting against
 the experimentally observed value of the mean life of neutron
 decay to proton. As the confined radius of the neutron is not
 precisely known, we use the well-known weak decay of neutron i,e,
  n$\rightarrow$p $\overline{e}$ $\overline{\nu}$. The beta decay
  coupling constant obtained from Fermi theory using the decay
  rate  $\Gamma_{fi}$ $\simeq$ 4[{2$\pi$}]$^{-3}$ $G^{2}$
  $m_{e}^{2}$. With the observed neutron life time =
  0.93*$10^{3}$ sec, we have obtained the $[{G m_{e}}]^{2}$ and $[{G
  m_{p}}]^{2}$ as 0.29x$10^{-11}$ and 1.10x$10^{-5}$
  respectively. Now the neutron and the proton are subject to strong interaction. We allow the possible strong interaction
  corrections i,e, pion correction introducing a form factor F,
  an invariant function of the proton and the neutron momenta so that the matrix element can be
  recast
  as F.[$\overline{u_{p}}$ $\gamma_{\mu}$ $u_{n}$] where symbols have
  their usual meanings. Hence the decay amplitude $\Gamma_{fi}$ =
  [(-16G$m_{p}m_{n}m_{e}m_{\nu})^\frac{1}{2}$
  ($\overline{u_{p}}\gamma_{\mu}u_{n}$)($\overline{u_{e}}\gamma_{\mu}u_{\nu}$) gets
  modified with this assumption.
  Now since the change in momentum involved
  is very small compared to the hadron masses, we do not expect
  much variation in F$(q^{2})$.  We assume $|q^{2}|$ =
  $(m_{n}-m_{p})^{2}$ = $\Delta^{2}$ which is almost constant. So
  when strong interaction is taken into account we may assume,
  \begin{equation}
  F. [G m_{p}]^{2}= 1.10x10^{-5}
  \end{equation}
   Now the form factor can be expressed as:
   \begin{equation}
   F(q^{2}) = \int e^{iq.r}|\psi(r)^{2}|
   \end{equation}
   In order to make some allowance of the strong interaction
   effect like pionic correction we assume that when two quarks
   come closer they interact strongly to yield pionic correction
   to the decay rate. Poggio et al [12] have pointed out that it
   would be good approximation that the maximum contribution would
   result from the ud di-quark region i,e, in the region r =
   $\frac{1}{2\mu}$, where $2\mu$ is the reduced mass of the ud
   diquark. For the small values of the $q^{2}$ we may recast the
   expression (5) as:
   \begin{equation}
   F(q^{2}) = \frac{4}{3}\pi r_{0}^{3}|\psi(r=\frac{1}{2\mu})|^{2}
   \end{equation}
   With $|\psi(r)|^2$ as input from (1)we obtain,
   \begin{equation}
   F = \frac{105}{16}(1-\frac{1}{m_{q}r_{0}}) ^{\frac{3}{2}}
   \end{equation}
   From (4) and (7) we estimated $r_{n}$ , the radius of the
   neutron as  0.80 fm using $m_{u}$ = $m_{d}$ = 360 MeV. We would
   use this value of neutron radius in our subsequent analysis.

     For proton we use
     the radius as 0.856fm. It may be mentioned here that Hefter [13]
     has applied the Non-Linear Schrodinger Equation (NLSE) [14]with
     an inhomogeneous term to the atomic physics and nuclear
     physics. The typical length of the system as described by NLSE is found to be related
     to the non-linear term of the equation. The objects are
     classified to be micro, meso and macro objects according to
     the characteristics the gausson solution. The solution involves a typical length which in turn related to the
     centre of mass gausson solution of  a spherical object with a
     typical radius  R(say). It has been pointed out
     that the properties of meso object are different from the
     other two types of objects. They have suggested that the nucleons and the $\alpha$ particles are subjected
     to the meso object.
      Hefter [13]has argued that the non linearities are related to the compressibility of
     the object. From the compressibility $\kappa$ of the proton
     it has been suggested that the proton belongs to a meso object
     with the one of the feature that its charge radius is
     equal to the radius parameter of the proton. They have indicated that the
      proton radius is $r_{p}$ = 0.865fm which
      equals to the most accepted
     value of the charge radius of proton. We have used
     this value of proton radius parameter in our subsequent
     analysis. Hefter [13] has mentioned that
     the maximum value
     of the nucleon matter radius should be 1.2fm from the study of the
     compressibility. we have also used $r_{p}$ = 1.2fm in some of
     the calculations.

   $\bf {The}$ $\bf{Structure}$ $\bf{function:}$

     The free nucleon $F_{2}$ structure function in the
     non-linear limit runs as:
     \begin {equation}
     F(x)=
     \frac{M}{8\pi^{2}}\int_{k_{min}}^{\infty}|\psi(k)|^{2}dk^{2}
     \end{equation}
     where M is the mass of the nucleon and $k_{min}$ = M $|x-
     \frac{1}{3}|$.
     With $\psi(k)$ as an input in (8) the structure function F(x) for
     the proton and neutron has been estimated using the relevant radius parameter as stated before. The results have
     been displayed in Fig-1 and in Fig-2.The difference between the structure functions has
     been estimated and results are displayed in the Fig-3. The ratio of the
     neutron proton structure functions has been displayed in the
     Fig-4. The ratios are estimated
     radius parameter of the proton as 0.856fm and 1.2fm.

     The Gottfried Sum rule runs as:
     \begin{equation}
     \int_{0}^{1} [\frac{F_{2}^{p}(x)- F_{2}^{n}(x)}{x}]dx = \frac{1}{3}
     \end{equation}
      The deviation of GT sum rule is a problem worth pursuing. It has been suggested that
       the deviation from the GT sum rule
     manifests the asymmetry between the $\overline{u}$ and
     $\overline{d}$ sea. We have estimated the GT sum rule in the
     context of our formalism with $r_{p}$= 0.865fm and $r_{n}$= 0.8fm and obtained the value as 0.240
     which agrees closely with corresponding result by NMC [2] which
     runs as $\pm$ 0.024 $\pm$0.034 $\pm$0.021. The ratio of the
     structure functions $\frac{F_{2}^{n}}{F_{2}^{p}}$ as x $\rightarrow$
      1 is estimated as 0.73 for aforesaid values of the neutron and
     proton radius. It may be mentioned that for $r_{p}$ = 1.2fm and $r_{n}$= 0.80fm we have
     obtained the ratio as 0.224.
     The ratio of the second moments  $\frac{M_{n}^{4}}{M_{p}^{4}}$
     has been obtained as 0.532 with the input of neutron and
     proton radius as 0.80fm and 0.865fm respectively which agrees
     well with the estimation of Osipenko et al [6].

     $\bf Results$ and $\bf Discussions$

       In the present work we have investigated the structure
       function properties of the neutron and the proton. The model we
       have used in the current investigation enable us to
       investigate the structure functions using only the size
       parameter of the respective nucleons.The results are presented in the
       range 0.1 $\langle$ x $\langle$ 0.5. Signal et al [15] have investigated the
       structure function for nucleon in the
       two dimensional MIT bag model with the cavity radius 1fm. $F_{p}$- $F_{n}$ estimated in
       the present work yields good agreement with the NMC results in the
       aforesaid region of x.
        Deviation from the GT
       sum rule estimated is found to be in good agreement with
       the NMC [2]results. Cheng et al [9] has indicated that a
       significant contribution to the deviation may come from the
       strange quark sea contribution of the nucleon. Broadhurst
       et al [8] have estimated GT sum rule $I_{g}$ as 0.219 to
       0.178. They have shown that the deviation persists even at
       the large $N_{c}$ limit. The ratio of the neutron proton
       structure functions estimated in the current work
       shows reasonable agreement with the work of Zamani et
       al [4] in which they have investigated the ratio in the light cone
        frame where a dressed nucleon is supposed to be
        superposition of the bare nucleon and a virtual light
        cone Fock state of baryon meson pairs. It may be mentioned that the ratio of
        the structure function is sensitive to the u and d quark
        distribution, particularly in the large x limit the ratio
        is sensitive to the valance distribution of u and d quark.
        With the spin -flavour symmetry, the SU(6) predicts the ratio as $\frac{2}{3}$ [16]. The
        $\frac{F_{2}^{n}}{F_{2}^{p}}$ in the large x limit (x
        $\rightarrow$1) in shell formalism has been estimated to be $\frac{1}{4}$ [17] whereas the helicity
         conservation in perturbative QCD
        with unperturbed spin flavour symmetric wave function
        yields the asymptotic value of the ratio as
        $\frac{3}{7}$ [18 ]. Osipenko et al [7 ] have  investigated the ratio in resonance region
         and have extracted
        the ratio as
        $\frac{F_{2}^{n}}{F_{2}^{p}}$ = 0.34 $\pm$ 0.12 $\pm$ 0.13 for x= 0.7. Melnitchouk
        et al [19 ] have extracted $\frac{F_{2}^{n}}{F_{2}^{p}}$ from the deuteron data and found that the result
        is consistent with the QCD prediction. Farrar et al [5]
        have
        investigated the problem in the  coloured quark and
        vector gluon model of hadron and observed that the result
        is in agreement with the QCD prediction. The result we
        obtain for the ratio is 0.73 for proton radius 0.865fm whereas
         with $r_{p}$ = 1.2fm we get the ratio as 0.22 close to the QCD prediction.

The radius parameter is a very important quantity which is yet to
be determined precisely.
             In the present work we have tried to investigate the
            properties of the nucleon structure function using only
            the radius parameter or in other words we have tried to investigate the value of the radius parameter of the
            proton and neutron through the investigation of the
            structure functions of the respective nucleons. The $r_{p}$ =.865fm and $r_{n}$ =.80fm are
            observed to be doing well in reproducing the structure
            function properties. The proton thus may has the possibility of
             behaving like a meso object. We have also used the
            radius parameter of the proton  as 1.2 fm which
            Hefter [13] has pointed out as the maximum value of the
            nucleon matter radius and obtained the good agreement
            with the experimental results.
            It is
            interesting to note that  deviation of GT sum rule which arises due
            to the light quark sea asymmetry has been found to
             be manifested by the spacial extension of the proton and
             neutron. The values of the structure functions obtained in the current work are found
             to be reasonably well whereas the GT sum rule and higher moments shows very good agreement.
              It may be mentioned that at
             large x symmetric sea vanishes more rapidly which may
             affect the experimental data. The results which are
             reproduced fitting with only one parameter is found to be
             very encouraging. It can be said that geometry plays an important role in exhibiting the structure function
             properties. It is pertinent to point out here that
             the the statistical model predicted the
             hadron as a fractal object with fractal dimension
             $\frac{9}{2}$ [11] and in the context of the model the scaling violation has been
             attributed to the fractal behaviour of the hadron.
             From the present study of the structure function using the model, it may not be irrelevant
             to emphasize that the
             violation of GT sum rule may be attributed to the
             fractal behavior of the nucleons. However further
             investigation would be done in our future works
             particularly considering the strange quark sea
             contribution to the GT sum rule.
             \newpage

Acknowledgements: Authors are thankful to T.L.Remadevi for useful
discussions.Authors are also thankful
             to Department of Science and Technology(DST), New
             Delhi, for financial assistance.

\newpage
{\bf References} \vskip .2in
\noindent
 [1] K. Hagiwara et al., Phys Rev. {\bf D 66} (2002) 010001
 ;P.V.Pobylitsa et al., Phys. Rev. {\bf D 59} (1999) 034024 ; D.Diakonov
 et al.,Nucl.Phys.{\bf B 480} (1996) 341.

\noindent [2] P. Amaudruz et al., Phys.Lett. {\bf B 295} (1992)
159.

\noindent [3] P. Amaudruz et al., Phys.Rev.Lett. {\bf 66} (1991)
2712.

\noindent [4] F. Zamini., Phy.Rev. {\bf C 58} (1998) 3641.

\noindent [5] G.R.Farrar et al., Nucl.Phys. {\bf B 480} (1996) 341
.

\noindent [6] M. Osipenko et al., Nucl.Phys. {\bf A 766} (2006)
142.

\noindent [7] M. Osipenko et al., Phys.Rev. {\bf D 67} (2003)
092001.

\noindent [8] D.J.Broadhurst et al., Phys.Lett. {\bf B 590} (2004)
76 .

\noindent [9] T.P.Cheng et al., Phys.Rev.Lett. {\bf 74} (1995)
2872.

\noindent [10] A.Bhattacharya et al., Prog.Theor.Phys. {\bf 77}
(1987) 16, Eur.Phys.J.C. {\bf 2}(1998) 671 , Int.J.Mod.Phys. {\bf
A 16}(2001) 201; S.N.Banerjee et al., Ann.Phys.(N.Y.) (1983)150.

\noindent [11] A.Bhattacharya et al., Nucl. Phys. {\bf A 728}
(2007) 292C ;  S.N.Banerjee et al., Phys.Lett. {\bf B 644} (2007)
45, Int.J.Mod.Phys. {\bf 17} (2002)4939.

\noindent [12] E.L.Poggio et al., Phys.Rev. {\bf D 20} (1979)1175
.

\noindent [13] E.F.Hefter., Phys.Rev. {\bf A 32} (1985)1201.

\noindent [14] I. Bialynicki-Birla et al.,
 Ann.Phys.(N.Y.)(1993)150.

\noindent [15] A.I.Signal et al., Phys.Lett. {\bf B 211}, (1988)
481.

 \noindent [16] F.E.Close., An introduction to quarks
and partons(Academic Press,New York,1979).

\noindent [17] J.J.Aubert et al., Nucl.Phys. {\bf B 293}
(1987)740; A.Bodek et al., Phys.Rev. {\bf D 20}(1979) 1471.

\noindent [18] S.J.Brodsky et al., Nucl.Phys. {\bf B 441}
(1995)197.

\noindent [19] W.Melnitchouk et al., Phys.Lett.{\bf B 377} (1996)
11.

\newpage
Figure captions:

1. Fig 1. The proton structure function with $r_{p}$ = 0.865fm.

2. Fig 2. The Neutron structure function with $r_{n}$ = 0.80fm.

3. Fig 3. The difference in proton neutron structure function
   $F_{2}^{p}$(x)- $F(x)_{2}^{n}$(x) with $r_{p}$ = 0.865fm and $r_{n}$ = 0.80fm.

4. Fig 4. The ratio of the neutron proton structure function
          $\frac{F_{2}^{n}}{F_{2}^{p}}$ for
          $r_{p}$ = 0.865fm. and 1.2fm whereas $r_{n}$ = 0.80fm.

\end{document}